\documentclass[12pt]{article}
\usepackage{amsfonts}
\usepackage{latexsym,amsmath}
\usepackage{graphicx}
\usepackage{placeins}
\usepackage{slashbox}
\usepackage{amssymb,array}
\makeatletter \oddsidemargin 0in \evensidemargin 0in \textwidth 16cm
\newcommand{\es}{\end{subsection}}

\begin{document}
 \pagestyle{empty}


\title{\large Cryptanalysis of Multi-Server Authenticated Key Agreement Scheme Based on Trust Computing Using Smart Cards and Biometrics}

\author{\small Dheerendra Mishra\thanks{E-mail:~{dheerendra@maths.iitkgp.ernet.in} }\\
\small Department of Mathematics,\\
\small Indian Institute of Technology Kharagpur,\\
\small  Kharagpur 721302, India\\}

\date{}
 \maketitle

\begin{abstract}
Advancement in communication technology provides a scalable platform for various services where a remote user can access the server from anywhere without moving from its place. It has provided a unique opportunity for online services, such that the user need not physically present at the service center. These services adopt authentication and key agreement protocols to ensure authorized and secure access to resources. Most of the authentication schemes support single server environment where the user has to register with each server. If a user wishes to access multiple application servers, he requires to register with each of the servers. Although multi-server authentication schemes introduced a scalable platform such that a user can interact with any server using single registration. Recently, Chuang and Chen proposed an efficient multi-server authenticated key agreement scheme based on smart cards along with password and biometrics. This is a lightweight authentication scheme which requires the computation of only hash function.  In this article, we present a brief review of Chuang and Chen's scheme. We analyze Chuang and Chen's scheme and identify that their scheme does not resist stolen smart card attack which causes the user's impersonation attack, server spoofing attack and man-in-the middle attack. Additionally, we show that their scheme has a weak key agreement protocol, which does not ensure forward secrecy. 

\end{abstract}

\noindent\textbf{keywords:} {Network Security; Multi-server communication; Smart card; Biometric based authentication; Anonymity.}

\section{Introduction}\label{intro}
The advances in communication technology  are enhancing the quality of online services. As a result, the Internet is emerging a scalable platform for various services where users and service providers are realizing its importance. This provides a unique opportunity to the users, such that they can access the remote servers at anytime and from anywhere. However, the user interacts with the server via public channel where an adversary is considered to be enough powerful that he can control over the public channel, such that he can eavesdrop, intercept, modify, delete, and replay the transmitted message via public channel. This causes a serious threat to the data security and integrity. At the same time, the authenticated  key agreement protocols provide secure and authorized communication between remote entities.
The smart card based authentication protocols are designed and developed to provide authorized and secure communication between the remote user and the server.

Most of the existing authentication protocol only supported a single server environment~\cite{tsai2008efficient}. However,  a user may wish to access multiple application servers at  the same time for various of kinds of application. Therefore, if the authentication scheme does not support the multi-server environment, the user has performed the registration on each of the servers. This makes the system very complex from the user point of view and a user may leave the system.

Most of the existing schemes that support multi server environment are password based~\cite{ pippal2013robust}. The password based authentication schemes provide two-factor remote user authentication while biometrics based user authentication schemes provide three-factor authentication. Moreover, uniqueness property of  biometric increases its application in authentication protocols. Therefore, biometric-based remote user authentication schemes have attracted significant research attention. The Biometric based schemes  have the advantages of biometric keys (fingerprint, face, iris, hand geometry and palm-print, etc.), which are as follows:\\(1) Biometric keys do not need to remember.\\ (2) Biometric keys cannot be easily guess.\\ (3) Biometric keys maintain the uniqueness property.

In 2010, Yang and Yang~\cite{yang2010biometric} proposed biometric-based multi-server authentication schemes. However, there scheme computational cost is high, as it needs to perform exponential operations.  In 2011, Yoon and Yoo~\cite{yoon2011robust} also presented a biometric-based authenticated  key agreement scheme for multi-server environment. Although He~\cite{he2011security} pointed out the vulnerabilities of Yoon and Yoo's scheme to insider attacks, masquerade attacks and loss of smart card attacks. Moreover, both the schemes fail to protect user's privacy.  Recently, Chuang and Chen~\cite{chuang2013anonymous} proposed an anonymous multi-server authenticated key agreement scheme based on smart cards along with password and biometrics. Their scheme provides an efficient solution for multi-server environment, where a user interact with any server using single registration.
In this article, we present a brief review of Chuang and Chen's scheme. We also present cryptanalysis of Chuang and Chen's scheme and show that their scheme does not resist stolen smart card attack which causes the user's impersonation attack, server spoofing attack and man-in-the middle attack. Additionally, we saw that their scheme has a weak key agreement protocol, which does not ensure forward secrecy.

The rest of the paper is organized as follows: Section \ref{review} presents the brief review of Chuang and Chen's scheme.   Section \ref{crypt} points out the weaknesses of Chuang and Chen's scheme. Finally, the conclusion is drawn in Section \ref{con}.

\section{Review of Chuang and Chen's Scheme}\label{review}

Recently, Chuang and Chen~\cite{chuang2013anonymous} proposed an anonymous authentication scheme using biometric-based smart card. 
 In this section, we will briefly discuss the  Chuang and Chen's scheme, in which we try to use the same terminology as presented in their article. 

\subsection{Server registration phase}
The application server sends a registration request to the registration center if he wishes to be become authorized server in the system. Then, registration center authorized the server and provides the key $PSK$ to the server using Key Exchange Protocol (IKEv2)~\cite{kaufman2005internet}. Upon receiving the secret key $PSK$, the authorized server uses this key to authorize the legitimate user.

\subsection{Registration Phase}
\begin{description}

\item[\bf Step 1.] The user $U$  computes $h(PW_i\oplus BIO_i)$ and sends his registration request with $ID_i$ and $h(PW_i\oplus BIO_i)$ to the registration center via secure channel.
\item[\bf Step 2.]  The registration center computes $A_i = h(ID_i||x), B_i = h(A_i) =  h^2(ID_i||x)$, $C_i = h(PW_i\oplus BIO_i)\oplus B_i$ and $D_i = PSK\oplus A_i$.

\item[\bf Step 3] Registration center personalizes the user's smart card by including the parameters $\{ID_i, B_i, C_i, D_i, h(\cdot)$ and provides the personalized smart card to the user via a secure channel.

\end{description}

\subsection{Login Phase}

To start the login session, the user inserts his smart card into the card reader and inputs his identity $ID_i$ and password $PW_i$, and imprints his biometric information $BIO_i$ at the sensor. Upon receiving the input, the smart card executes the login session as follows:

\begin{description}
\item[\bf Step 1.] Verify $ID_i$ and $B_i \overset{?}{=}~ h(PW_i\oplus BIO_i)\oplus C_i$. If the verification succeeds, it executes the next step.

\item[\bf Step 2.] Generate a random number $N_1$ and compute $M_1 = h(B_i)\oplus N_1$, $AID_i = h(N_1)\oplus ID_i$ and $M_2 = h(N_1||AID_i||D_i)$.

 \end{description}

\subsection{Authentication Phase}
\begin{description}
\item[\bf Step 1.] The smart card sends the  authentication request with the message $<AID_i, M_1, M_2, D_i>$ to the server.

\item[\bf Step 2.] Upon receiving the message $<AID_i, M_1, M_2, D_i>$, server $S$ uses its pre-shared key $PSK$ and achieves $A_i = D_i\oplus PSK$. The server also retrieves $N_1 =  h(B_i)\oplus M_1$ and $ID_i = AID_i\oplus h(N_1)$.

\item[\bf Step 3.] The server verifies  $M_2 \overset{?}{=} ~h(N_1||AID_i||D_i)$. If verification holds, the server generates a random number $N_2$ and computes the session key $SK_{ij} = h(N_1||N_2)$.
\item[\bf Step 4.] The server computes $M_3 = N_2 \oplus h^2(N_1)$ and $M_4 = h(SID_j||N_2)$ and responds with the message $<SID_j, M_3, M_4>$ to the smart card.

\item[\bf Step 5.] Upon receiving the message $<SID_j, M_3, M_4>$, the smart card retrieve the value $N_2 = M_3  \oplus h^2(N_1)$. Then, it verifies $M_4 \overset{?}{=} ~h(SID_j||N_2)$. If verification holds then computes the session key $SK_{ij} = h(N_1||N_2)$.

\item[\bf Step 6.] The smart card computes $M_5 = SK_{ij} \oplus h(N_2)$ and sends $M_5$ to the server.

 \item[\bf Step 7.] Upon receiving  $<M_5>$, the server verifies $h(N_2) \overset{?}{=}~ M_5\oplus SK_{ij}$. If the varication holds, the mutual authentication completes.

\end{description}

\subsection{Password change phase}

The legal user can change his password without the help of server as follows:

\begin{description}
\item[\bf Step 1.]  A user inputs his identity $ID_i$ and $PW_i$, and imprints his biometric $BIO_i$ at the sensor.

  \item[\bf Step 2.] The smart card verifies $ID_i$ and  $B_i \overset{?}{=}~h(PW_i\oplus BIO_i) \oplus C_i$. If the verification does not succeeds, the smart card rejects the request. Otherwise, the user can enter a new password $PW_i^*$.

\item[\bf Step 3. ] The smart card computes $C_i^* = C_i\oplus h(PW_i\oplus BIO_i) \oplus h(PW_i^*\oplus BIO_i)$ and replace $C_i$ with $C_i^*$.

\end{description}

\section{Cryptanalysis of Chuang and Chen's Scheme}\label{crypt}
In this section, we analyze Chuang and Chen's scheme and demonstrate some of the attacks. 
%
%

\subsection{Stolen smart card attack}
An efficient biometric based multi-server authentication protocol must not allow an adversary to misuse user's stolen smart card to login to the server or to compute established session keys without knowing the user's biometric and password. Here, we show that  Chuang and Chen's scheme fails to resist stolen smart card, such that an adversary can achieve user's long term secret key and can easily login to the server as a legitimate user using stolen smart card. Additionally, an adversary can achieve previously established session keys. This creates the data security and integrity threat as user and server protect their confidential data during data transmission using the session key and if session key is compromised, an adversary can achieve all the data that have been transferred between the user and server. The stolen smart card attack executes on Chuang and Chen's scheme as follows:

\begin{itemize}
  \item An adversary can achieve the stored parameters $\{ID_i ,B_i, C_i$ and $D_i$ from the smart card using existing techniques, such as power analysis attack, differential attack etc. 
  \item An adversary can intercept and record all the previously transmitted message $M_1 = h(B_i)\oplus N_i$ and $M_3 = N_2 \oplus h^2(N1)$.
  \item The adversary can achieve the values $N_1 = M_1\oplus h(B_i)$ and $N_2 = M_3 \oplus h^2(N_1)$.

 \item Using the values $N_1$ and $N_2$, the adversary achieve the session key $SK_{ij}$ as  $SK_{ij} = h(N_1||N_2)$.
\end{itemize}

It is clear from the above discussion that an adversary can achieve all the previously established session key using the stolen smart card. Using the compromised session key, an adversary can achieve all the confidential data that is transferred between user and server as exchanged data is being protected by established session key and an adversary can eavesdrop and intercept all the transmitted message between  user and server.

\subsubsection{User's Impersonation attack}\label{user's-impersonation}
An adversary can also successfully login to the server using the stolen smart card without having the user's biometric imprint and password as follows:

\begin{itemize}
 \item An adversary can achieve the stored parameters $ID_i ,B_i$ and $D_i$ from the smart card.
  \item  The adversary generates a random number $N_E$ and computes $M_{1E} = h(B_i)\oplus N_E$, $AID_E = h(N_E)\oplus ID_i$ and $M_{2E} = h(N_E||AID_E||D_i)$. He  masquerades as a legitimate user $U$ and sends the message $<AID_E, M_{1E}, M_{2E}, D_i>$ to the server.
  \item  Upon receiving the message $<AID_i, M_{1E}, M_{2E}, D_i>$, the server achieves  $N_E = h(B_i)\oplus M_{1E}$ and $ID_i = AID_E\oplus h(N_E)$. Then, the server verifies the condition  $M_{2E} \overset{?}{=} ~h(N_E||AID_E||D_i)$. The verification holds as $M_{2E} = h(N_E||AID_E||D_i)$.  When verification holds, the server generates a random number $N_2$ and computes the session key $SK_{ij}' = h(N_E||N_2)$.

\item The server computes $M_{3E} = N_2 \oplus h^2(N_E)$ and $M_4 = h(SID_j||N2)$ and sends the message $<SID_j, M_{3E}, M_4>$ to the user $U$.

\item The adversary intercepts the message $<SID_j, M_{3E}, M_4>$ and retrieves the value $N_2 = M_{3E} \oplus h^2(N_E)$ and $SK_{ij}^* = h(N_E||N_2)$. He responds with the message $M_5^* = SK_{ij}^* \oplus h(N_2)$ to the server.

\item Upon receiving  $M_5^*$, the server verifies $h(N_2) \overset{?}{=}~ M_5^*\oplus SK_{ij}'$. The verification holds as  $SK_{ij}' = SK_{ij}^* = h(N_E||N_2)$. This shows that the adversary successfully  masquerade as a legitimate user using the stolen smart card.
\end{itemize}



%


\subsection{Server spoofing attack}
 Chuang and Chen's scheme is vulnerable to the server spoofing attack, $i. e. $, an adversary can impersonate the server to the user. The detailed description is as follows:

\begin{itemize}
\item The adversary eavesdrops communication between users' and smart card. He intercepts the server's response message $<SID_j, M_3, M_4>$ and achieves server's identity $SID_j$.

  \item When the smart card transmits the  authentication request with the message $<AID_i, M_1, M_2, D_i>$ to the server $S$ via public channel, the adversary intercepts the message. Then, the adversary responds with authorized message using the achieved parameters $ID_i$ and $B_i$ from the stolen smart card as follows:\\
                    1.) Compute $N_1 = M_1\oplus h(B_i)$.\\
                   2.) Generate a random number $N_E$.\\
                  3.) Compute $M_{3E} = N_E\oplus h^2(N_1)$ and $M_{4E} = h(SID_j||N2)$, and responds with the message $<SID_j, M_{3E}, M_{4E}>$ to the smart card.

  \item When the smart card retrieves $N_{2E} = M_{3E}\oplus h^2(N_1)$ and verifies $M_{4E} \overset{?}{=} ~h(SID_j||N_{2E})$. The verification holds as $M_{4E} = h(SID_j||N2)$.
  \end{itemize}
This shows that an adversary can successfully impersonate as a server.

\subsection{Man-in-the middle attack}
 Chuang and Chen's scheme is vulnerable to the man-in-the middle attack. The justification is as follows:
\begin{itemize}
  \item   When the smart card sends the  message $<AID_i, M_1, M_2, D_i>$ to the server, the adversary intercepts the message and performs the following  steps:

  i.) Compute $N_1 = M_1\oplus h(B_i)$.\\ ii.)
        Generate a random number $N_E$.\\ iii.)
      Compute $M_{1E} = h(B_i)\oplus N_E$, $AID_E = h(N_E)\oplus ID_i$ and $M_{2E} = h(N_E||AID_E||D_i)$ then sends the message $<AID_E, M_{1E}, M_{2E}, D_i>$ to the server.\\ iv.)
   Compute $M_{3E} = N_E \oplus h^2(N_1)$ and $M_{4E} = h(SID_j||N_E)$ then sends the message $<SID_j, M_{3E}, M_{4E}>$ to the smart card.

  \item When the server retrieves $N_E = h(B_i)\oplus M_{1E}$ and $ID_i = AID_i\oplus h(N_E)$ and verifies  $M_2 \overset{?}{=} ~h(N_1||AID_i||D_i)$. The verification holds as $M_{4E} = h(SID_j||N_E)$. Then the server computes the session key $SK_{Ej} = h(N_E||N_2)$.
   \item When the server responds with the message $<SID_j, M_3', M_4'>$, where $M_3' = N_2 \oplus h^2(N_E)$ and $M_4' = h(SID_j||N_2)$. The adversary intercepts the message and executes the following steps:

    a.) Compute $N_2 = M_3'\oplus h^2(N_E)$ and $SK_{Ej} = h(N_E||N_2)$.\\
     b.) Send the message $SK_{Ej} \oplus h(N_2)$ to the server.

  \item When the server verifies $h(N_2) \overset{?}{=}~ SK_{Ej} \oplus h(N_2)\oplus SK_{Ej}$. The verification holds.

\item When the smart card retrieves the value $N_E = M_{3E}  \oplus h^2(N_1)$ and verifies $M_{4E} \overset{?}{=} ~h(SID_j||N_E)$. The verification holds as $M_{4E} = h(SID_j||N_E)$. The smart card computes the session key $SK_ {I} = h(N_1||N_E)$ as the verification holds.

\item When the smart card sends $SK_{iE} \oplus h(N_E)$ to the server, the adversary intercept the message.

\item The adversary can compute the session key with user and server $SK_{iE} = h(N_1||N_E)$ and $SK_{Ej} = h(N_E||N_2)$, respectively.
\end{itemize}
It is clear from the above discussion that an adversary can make independent connections with both the user and server where the user and serve believe that they are communicating directly with each other. Moreover, the user and server compute the keys, such that $SK_{iE} \neq SK_{Ej}$. Although the adversary computes both the session keys $SK_{iE}$ and $SK_{Ej}$.

\subsection{Forward secrecy}
Chuang and Chen's scheme does not achieve forward secrecy as the adversary can compute established session key  with the user's compromised long-term secret key $A_i$ as follows:

\begin{itemize}
  \item Achieved user's and server's transmitted message $<AID_i, M_1, M_2, D_i>$ and $<SID_j, M_3, M_4>$ via public channel, respectively.
  \item Compute $B_i = h(A_i)$ and retrieve $N_1 = M_1\oplus h(B_i)$ and then $N_2 = M_3 = N_2 \oplus h^2(N_1)$.
  \item  Compute the session key $SK_{ij} = h(N_1||N_2)$.

\end{itemize}



\section{Conclusion and future scope}\label{con}
 We have shown that an adversary can successfully perform the stolen smart card attack which causes the user's impersonation attack, server spoofing attack and man-in-the middle attack. We have also demonstrated that Chuang and Chen's scheme does not ensure forward secrecy. 

%

\begin{thebibliography}{1}
\providecommand{\url}[1]{#1}
\csname url@samestyle\endcsname
\providecommand{\newblock}{\relax}
\providecommand{\bibinfo}[2]{#2}
\providecommand{\BIBentrySTDinterwordspacing}{\spaceskip=0pt\relax}
\providecommand{\BIBentryALTinterwordstretchfactor}{4}
\providecommand{\BIBentryALTinterwordspacing}{\spaceskip=\fontdimen2\font plus
\BIBentryALTinterwordstretchfactor\fontdimen3\font minus
  \fontdimen4\font\relax}
\providecommand{\BIBforeignlanguage}[2]{{%
\expandafter\ifx\csname l@#1\endcsname\relax
\typeout{** WARNING: IEEEtran.bst: No hyphenation pattern has been}%
\typeout{** loaded for the language `#1'. Using the pattern for}%
\typeout{** the default language instead.}%
\else
\language=\csname l@#1\endcsname
\fi
#2}}
\providecommand{\BIBdecl}{\relax}
\BIBdecl

\bibitem{tsai2008efficient}
J.-L. Tsai, ``Efficient multi-server authentication scheme based on one-way
  hash function without verification table,'' \emph{Computers \& Security},
  vol.~27, no.~3, pp. 115--121, 2008.

\bibitem{pippal2013robust}
R.~S. Pippal, C.~Jaidhar, and S.~Tapaswi, ``Robust smart card authentication
  scheme for multi-server architecture,'' \emph{Wireless Personal
  Communications}, vol. {72}, no. {1}, pp. 1--17, 2013, doi={10.1007/s11277-013-1039-6}.

\bibitem{yang2010biometric}
D.~Yang and B.~Yang, ``A biometric password-based multi-server authentication
  scheme with smart card,'' in \emph{International Conference on Computer
  Design and Applications (ICCDA 2010)},  pp. 554--559, 2010.

\bibitem{yoon2011robust}
E.-J. Yoon and K.-Y. Yoo, ``Robust biometrics-based multi-server authentication
  with key agreement scheme for smart cards on elliptic curve cryptosystem,''
  \emph{The Journal of Supercomputing}, vol.~63, no.~1, pp. 235--255, 2011.

\bibitem{he2011security}
D.~He, ``Security flaws in a biometrics-based multi-server authentication with
  key agreement scheme.'' \emph{IACR Cryptology ePrint Archive}, vol. 2011, p.
  365, 2011.

\bibitem{chuang2013anonymous}
M.-C. Chuang and M.~C. Chen, ``An anonymous multi-server authenticated key
  agreement scheme based on trust computing using smart cards and biometrics,''
  \emph{Expert Systems with Applications}, 2013, doi = "http://dx.doi.org/10.1016/j.eswa.2013.08.040.

\bibitem{kaufman2005internet}
C.~Kaufman, ``Internet key exchange (ikev2) protocol,'' 2005.

\end{thebibliography}


\end{document}